\documentclass[12pt]{article}
\usepackage{amssymb,amsfonts,a4}

\newcommand{\GG}{G}
\newcommand{\tr}{\,\mbox{tr}}
\begin{document}
\title{From Yang--Mills Field to Solitons and back again}
\author{\bf L.D.~Faddeev}
\date{}
\maketitle
\begin{center}
{\it Steklov Mathematical Institute, St.Petersburg}\\
{\it Helsinki Institute of Physics, Helsinki}\\
\end{center}
\medskip

I am here in Erice for the first time in my life.
Of course I followed the activity of this famous School
during the last 35 years via its Proceedings, some of
which certainly influenced my scientific interests.
But now I am able to watch the real work of the School
and this is a great revelation. I am very grateful to
Professors ~Zichichi, 't~Hooft and ~Veneziano for inviting
me here.

When I got the invitation, I decided to come and use
the opportunity to speak in defence of Quantum Field
Theory. The fact that I was to be the last speaker is
certainly helpful for this goal. However, I found here
that there is nobody present at the School to argue
with. All my colleagues theorists apparently are in the
mood of supporting QFT, at least in the guise of QCD.
So it remains for me just to explain why I personally
like QFT. Needless to say that my arguments will be
significantly based on my own experience.

Quantum Field Theory is only a little younger than Quantum
Mechanics. The realization of a field as a system of
infinite number of oscillators was used to construct
the quantization of fields already in 1928 by fathers
of QM --- Heisenberg, Pauli, Jordan, Dirac. My senior
compatriot V.~Fock entered the field almost at the same
time. The first essential success of QFT, which at that
time was Quantum Electrodynamics, was the resolution
of the long standing contradiction in connection with
the origin of light. Both wave and corpuscular aspects
were united naturally in the quantum description of the free
electromagnetic filed: the quantized hamiltonian of the
Maxwell wave equations has the particle-like excitations ---
photons. The subsequent calculations of the basic processes
of interacting electrons and photons was a great success.
However the infinities which appeared in the radiative
corrections, became the uncircumventable obstacles for the
following development and interest to QED temporarily
declined in the end of 30-ties.

New great success of QED is connected with the understanding
of the radiative corrections prompted by new
experimental data during 1948--1953. Advent of the
manifest Lorentz invariant formulation and
renormalization by Tomonaga, Schwinger and Feynman
allowed to calculate the Lamb shifts of atomic levels
and anomalous magnetic moment of electron with great
precision and agreement with experiment. The general
renormalization procedure, developed by Dyson, Salam,
Bogolubov among others and based on Feynman
diagrammatic approach to perturbation expansion, became
a new paradigm of QFT. However, the (now looking) naive
generalization of QED to strong nuclear forces with
$ \pi $--meson field substituting the electromagnetic
field of QED, did not lead to satisfactory results.
Even worse was the realization that the charge
renormalization in QED leads to the increasing of the
interactions for high energies. The extremal position
in this respect was taken by Landau, who pronounced
the Hamiltonian method (i.e. all QFT) dead in 1956.
All this led to the second decline of QFT with
interests of the theoreticians shifted to nondynamical
S--matrix theory. In my country the QFT was virtually
forbidden from the end of 50-ties for at least ten
years. Fortunately for me I lived in Leningrad (now
St.Petersburg, where I continue to stay) and worked in
the Mathematical Institute; so I was not subject to
the prevailing ideological trends in Soviet theoretical
physics. So when in the middle of the 60-ties I decided to
become seriously involved in QFT, I had no
inhibitions. The search for a possible subject almost
immediately led me to Yang-Mills field. There was no
compelling physical reason to consider this field
being charged and massless. However the beautiful
geometric interpretation, which I learned from
Lichnerowics's book on the theory of connections
\cite{c1},
was irresistible. One more reason to do YM was the
influence of Feynman lecture of 1962 on quantization
of Einstein Gravitation Theory
\cite{c2}. The YM field
was used there as a toy model, and Feynman found that the
conventional diagrammatic rules of the perturbation theory do
not work.

For reasons not understandable to me Feynman did not return
to the fundamental source of his diagrammatic rules ---
namely, Feynman functional integral. Indeed, it is in
the functional integral formalism where the
specificity of the YM field can be seen most
transparent. The equivalence principle states, that
gauge equivalent fields are physically
indistinguishable, so in the functional integral over
all physical configurations one is to integrate over
classes of equivalent fields, choosing one
representative in each class. In QED the gauge orbit
\begin{equation}
 \label{f1}
A_{\mu} \to A_{\mu}^{\varphi} =
A_{\mu} + \partial_{\mu} \varphi \, ,
\end{equation}
parametrized by a real function
$ \varphi(x) $
on space--time, is linear and typical gauge
condition like
$ \partial_{\mu} A_{\mu} = 0 $,
serving for choice of representative, intersects all
orbits with a fixed angle. In the case of nonabelian
YM field the situation is different: the orbits
\begin{equation}
\label{f2}
A_{\mu} \to A_{\mu}^{g} =
g^{-1} A_{\mu} g + g^{-1} \partial_{\mu} g \, ,
\end{equation}
where
$ g(x) $
is a function on space--time with values in compact group
$ \GG $, are nonlinear and intersect the gauge surface
$ \partial_{\mu} A_{\mu} = 0 $
with angle depending on orbit. What measures this
angle is a determinant
\begin{equation}
\label{f3}
\Delta(A) = \mbox{det} \, \partial_{\mu} \nabla_{\mu} =
  \mbox{det} \left( \square \partial_{\mu}
  + \left[ A_{\mu},\, \cdot ~\right]
   \right)
\end{equation}
which
certainly is to be taken into account in the
functional integral.

This geometrical idea was formally realized by my
collaborator V.~Popov and me in the fall of 1966 and led to
the concrete proposal for the modification of the diagrammatic
rules for YM field. Due to the circumstances mentioned above,
it was not possible to publish a long paper on this
subject in any Soviet journal, so we used a very
new opportunity of relative liberalization to send a
short letter to Physics Letters
\cite{c3} and published more
extended version as a preprint
\cite{c4} of just organized Kiev
Institute of Theoretical Physics. Complete set of
Feynman diagramatic rules as well as interpretation
of additional action due to (\ref{f3}) in terms of
integral over fictitious fermions ( now called ghosts)
is contained in this preprint. Also the equivalence
of manifestly covariant gauge and Coulomb gauge, where only
physical degrees of freedom entered the functional intrgral,
was established in this preprint. It showed the unitarity of the
modified diagramatic rules. More general discussion of the
quantization of the constrained Hamiltonian systems (of which
the Yang-Mills field theory gives a prominent example) was
given by me in
\cite{f1}.

Later, when YM theory came to vogue, the English translation
of Kiev preprint was published by B.~Lee as a Fermilab
preprint. However, in the middle of the 60-ties our
results did not attract much attention. The evident
reason was the lack of viable physical applications.

The situation changed drastically with 't~Hooft's applications
of Yang--Mills in the model with spontaneous breaking of the
symmetry
\cite{c5}, which turns the charged components of
vector field into the massive ones. The relevance of this
model to the theory of weak interactions immediatly
attracted a lot of attention of physicists.

The subsequent development including the
dramatic history of the asymptotic freedom is
described in 't~Hooft's lecture at this School and in
recent talk of D.~Gross
\cite {g1}, so I
must not repeat it. As a result, QFT was vindicated and
returned to its well deserved position of the
fundamental base for the theory of elementary
particles. The Standard Model based on Yang--Mills
field stands until now as a satisfactory description
of all interactions but gravity.

In what follows I shall concentrate on some new features of
the Third Advent of QFT, which distinguishes it from two
previous periods.

The obvious drawback of pure YM theory --- the absence of
mass parameter in classical formulation --- becomes, in fact,
its asset. Indeed, the only parameter entering the theory is
dimensionless coupling constant
$ 1/g^{2} $,
standing in front of action very much like Planck constant
$ 1/\hbar $.
The asymptotic freedom can be reformulated in the
following
way: the regularization introduces the cut-off momentum
$ L $
of dimension of mass (for example,
$ L= 1/\Delta $, where
$ \Delta $
is lattice spacing) and with increase of
$ L $
the coupling constant is to go to zero. The explicit rule in
the simplest approximation looks like
\begin{equation}
\label{f4}
\ln \frac{L}{\mu} - \frac{c}{g^{2}} = \ln \frac{m}{\mu} \, ,
\end{equation}
where
$ \mu $
is an arbitrary scale and
$ c $
is a positive constant depending on the gauge group.
Two infinities in the limit $ L \to \infty
$, $ g^{2} \to 0 $ in LHS are to cancel each other
leaving finite RHS and introducing a new parameter $ m
$. The formula above can be rewritten as
\begin{equation}
\label{f5}
m = L e^{-c/g^{2}}
\end{equation} expressing $ m $ via $ g^{2}
$ and $ L $ more explicitly.

Thus, the divergence in quantum version of YM field becomes
the origin of appearance of mass, which substitutes the
classical dimensionless coupling constant. This striking
phenomenon was called ``dimensional transmutation'' by
S.~Coleman in his famous Erice lectures of 1973
\cite{c6}.

I believe that this is one of the most important lessons of
the new period of QFT: some infinities are not bad, on the
contrary they are necessary for the physical applicability of
QFT. At least one infinity is good. Of course the QCD
using pure YM field without Higgs field is based on
dimensional transmutation.

It is instructive to comment here, that there exists a simple
example in ordinary Quantum Mechanics where formula like
(\ref{f4}) is exact.

The hamiltonian
\begin{equation}
\label{f6}
H = - \frac{\partial^{2}}{\partial x^{2}}
  - \frac{\partial^{2}}{\partial y^{2}} -
 g^{2} \delta(x)\delta(y)
\end{equation}
for two-dimensional particle interacting with a point source
of strength
$ g^{2} $
is apparently scale invariant. However, one encounters a
divergence
which is circumvented by renormalization of
$ g^{2} $
exactly as in
(\ref{f4}),
thus trading it for dimensional parameter breaking scale invariance.

The view on importance of infinity for QFT advertized
above was stressed also by R.Jackiw in
\cite{c7}.

Another paradigm of old QFT states that one is to introduce
a separate field for each elementary particle. In view of
increasing a list of such particles a new mechanism for mass
spectrum allowing many particles to correspond to one or a few
fields is needed. Of course, the Standard Model is quite happy
with its 12 vector fields and 12 fermions; but any
possible reduction of the number of consistuents is
certainly welcome.

A new mechanism of rich particle spectrum was found in
quantum soliton theory. I believe, that in spite of
the fact that we know about solitons already 25 years,
their application to the theory of elementary
particles only begins.

The term soliton appeared in applied plasma physics and was
introduced by M.~Kruskal and N.~Zabuski to describe a solitary
wave solution of one-dimensional nonlinear evolution
equation, so called Korteweg-de-Vries (KdV) equation.
In spite of the fact that KdV equation is very far from
relativistic physics, it surfaced several times in HEP-TH
in connection with Conformal Field Theory (CFT) and
Matrix Models. However another equation admitting solitons
is more relevant for us now. It is the so--called
Sine--Gordon equation
\begin{equation}
\label{f7}
  \varphi_{tt} - \varphi_{xx} + \frac{m^{2}}{\beta}
   \sin\beta\varphi = 0 ~~~~,
\end{equation}
which describes massive scalar field
$ \varphi(x) $
in 1+1 dimensional space--time; parameter
$ \beta $
is a coupling constant.

The pure step
\begin{equation}
\label{f8}
\varphi_{0}(x,t) = \left\{
\begin{array}{cc}
0 & x < x_{0} \\
\frac{2\pi}{\beta} & x > x_{0}
\end{array}
\right.
\end{equation}
is a formal stationary solution for all
$ x \neq x_{0} $.
There exists also a regular stationary solution,
\begin{equation}
\label{f9}
 \varphi(x) = \frac{4}{\beta} \arctan e^{mx}
\end{equation}
which formally converges to
$ \varphi_{0} $ when
$ m \to \infty $.
Both solutions connect two adjacent vacua
$ \varphi = 0 $ at $ x = - \infty $
and
$ \varphi = 2\pi / \beta $ at $ x = \infty $;
the solution
$ \varphi(x) $
has a finite mass (energy at rest)
\begin{equation}
\label{f10}
 M_{S} = \frac{8m} {\beta^{2}}
\end{equation}
and the running solution
\begin{equation}
\label{f11}
 \varphi_{v} (x,t) = \varphi \left(\frac{x-vt}
    {\sqrt{1-v^{2}}}
    \right)
\end{equation}
evidently corresponds to a particle. Let us stress that
perturbative paradigm states that equation
(\ref{f7})
describes just a particle of mass
$ m $.
Equation
(\ref{f7})
admits also a family of periodic solutions (breathers),
which can be interpreted as bound states of two kinks with
mass
\begin{equation}
\label{f12}
 M_{B} = 2 M_{S} \sin \theta ~~~~,
\end{equation}
where
$ \theta $, $ 0 < \theta \leq \pi/2 $
characterises the internal periodic motion.

Quantization of this picture, performed quasiclassically by
L.~Takhtajan and me in 1973
\cite{c8}, showed that variable
$ \theta $
became quantized, so that breathers produced the particle
spectrum
\begin{equation}
\label{f13}
 m_{n} = \frac{2m}{\gamma} \sin \frac{\gamma n}{2} ~~~~,
n=1, \ldots , \left[\frac{\pi}{\gamma}\right] ~~;
~~\gamma = \frac{\beta^{2}}{8}~~~.
\end{equation}
Later R.~Dashen, B.~Hasslacher and A.~Neveu showed
\cite{c9}, that one loop correction leads to the finite
renormalization
\begin{equation}
 \label{f14}
  \gamma \to \gamma' =
\frac{\gamma} {\pi - \gamma}
\end{equation} which, in
particular, shows, that $ \gamma $ can not exceed $
\pi $. Another interpretation of this fact was proposed by
S.~Coleman
\cite{c10}, who found a duality between SG model
and Massive Thirring Model describing self
interacting massive fermion. Soliton corresponds to
basic fermion, whereas breather excitations are its
bound states. The most striking feature of this
duality is exchange of weak and strong interactions in
two descriptions of the same physical system. It
became prototype for many dualities discussed
subsequently in physical literature, beginning with
the paper of Montonen and Olive
\cite{c11}.

I have no time to discuss the development of Quantum Theory of
Solitons with which I was connected for a long time since the
beginning of the 70--ties. The achievements include the
invention of Quantum Groups with their numerous
applications both in Mathematics and CFT. A lot of
examples of 1+1 dimensional models were quantized
exactly, in particular the nonlinear
$ \sigma $--model, where
the dimensional transmutation formula \`a-la
(\ref{f5})
was derived rigorously.

In more realistic 3+1 dimensional space--time the success of
solitons was less impressive. However, for
$ SU(2) $
YM field Yang and Wu
\cite{c12} introduced a singular soliton
\begin{equation}
\label{f15}
A_{0}^{a} = 0 ; \quad A_{k}^{a} = \frac{\epsilon_{iak}x^{i}}{r^{2}}
\end{equation}
which is somehow an analogue of step--like soliton
$ \varphi_{0}(x,t) $
of SG model. Physically Yang--Wu soliton describes magnetic
monopole. Indeed, one can find a gauge transformation,
singular along the Dirac string, which will turn
(\ref{f15})
into the usual configuration of abelian magnetic monopole
(more on this later).

There exists no smoothed solution of YM equation, analogous
to kink soliton
$ \varphi(x,t) $
of the SG model. However, such smoothening appears for the
model including the Higgs field. It was shown independently by
't~Hooft
\cite{c13}
and A.~Polyakov
\cite{c14}
in 1974.
Polyakov coined the name "hedgehog" for such a soliton. The
monopole charge characterizes the following topological
situation: the Higgs field of the static solution acquires
different asymptotic values in different directions in
3--space
$ {\mathbb R}^{3} $
(thus the hedgehog) and defines the map of the sphere
$ {\mathbb S}^{2} $
of such directions into space of vacua moduli. In the case of
$ SU(2) $
group this space is also
$ {\mathbb S}^{2} $.
The map
\begin{equation}
\label{f16}
 {\mathbb S}^{2}_{\mbox{\scriptsize space boundary}} \to
 {\mathbb S}^{2}_{\mbox{\scriptsize moduli}}
\end{equation}
 has a topological degree
 --- number of times the base covers the target --- which
is exactly the magnetic charge of the monopole. Thus
topology produces charge, which conserves
irrespectively of equations of motion. This feature is one
more substantial addition to QFT brought in during the new
period.

It is appropriate to remark here, that in
his quest for nonquantum particles Einstein was
eager to get such a picture
\cite{c15}.
So topological charges in some sense realize the
Einstein's dream. However, in modern setting they stay
against quantum principles in no way. On the contrary,
topological charge enter quantum theory most
naturally, e.g., via functional integral.

There is another possibility for the topological
charges besides hedgehog-like configuration. It
corresponds to really localized solitons, for which
the boundary condition at infinity in space do not
depend on direction. In this way the space
$ {\mathbb R}^{3} $
itself is compactified by addition of the point at
infinity and topologically became a sphere
$ {\mathbb S}^{3} \, $.
The static field then maps this
$ {\mathbb S}^{3} $
into the target space. As soon as this latter is
compact, one can expect appearance of topological
numbers.

An explicit example is given by Skyrme model, in
which the target space is a compact group, so that the
field corresponds to nonlinear
$ \sigma $--model.
It is relevant to stress here, that Skyrme was the
pioneer of using particle--like solution in high
energy physics, his papers
\cite{c16},
appeared much before
the solution fashion of mid 70--ties, however, he did
not discuss quantization.

The field
$ g(x) $
of Skyrme model has values in compact group
$ \GG $.
The boundary condition is
\begin{equation}
  \label{f17}
  g \bigr|_{r \to \infty} = g_{0}~~~,
\end{equation}
where
$ g_{0} $  is a fixed element in
$ \GG $, i.e., just a
unity. The lagrangian is described as follows.
Introduce the current
\begin{equation}
  \label{f18}
   L_{\mu} = \partial_{\mu} g g^{-1}~~,
\end{equation}
which gets its values in the Lie algebra of $ \GG $.
Then the lagrangian is defined as
\begin{equation}
 \label{f19}
 {\cal L} = a \tr L_{\mu}^{2} + b \tr
 \left ( \left[ L_{\mu}, L_{\nu}\right] \right)^{2}
\end{equation}
where
$ a $  and
$ b $  are coupling constants,
$ a $
has dimension of
$ (\mbox{mass})^{2} $  and
$ b $  is dimensionless. The
topological current
\begin{equation}
 \label{f20}
  J_{\mu } = \epsilon_{\mu\nu\rho\sigma} \tr
 L_{\nu} \left[ L_{\rho}, L_{\sigma}
  \right]
\end{equation}
conserves
$ \partial_{\mu} J_{\mu} = 0$
irrespective of
equations of motion and topological charge
\begin{equation}
\label{f21}
 Q = \int J_{0} d^{3} x
\end{equation}
acquires integer values upon some normalization.

It is evident, that static Hamiltonian density,
corresponding to lagrangian
(\ref{f19})
is a sum of squares of
entities, product of which defines the density of
topological charge.
Thus the estimate
\begin{equation}
\label{f22}
 H_{static} \ge c | Q |
\end{equation}
is valid, showing the possibilities for nontrivial static
solitons in sectors with
$ Q \ne 0 $.
Indeed, the Skyrme
soliton for
$ \GG = SU(2) $ and
$ Q = 1 $ can be described rather
explicitly and higher sectors produce a very rich
family of particle--like solutions (see recent paper
\cite{c18}).

One of additions to 3+1 dimensional solutions in
mid 70--ties was my proposal to modify Skyrme model
using as a target the 2--sphere
$ {\mathbb S}^{2} $,
 so that the
field is a unit direction vector
\begin{equation}
  \label{f23}
  \vec{n} = (n^{1},n^{2},n^{3}) ~~~, n^{2} = 1~~~.
\end{equation}
The boundary condition is
\begin{equation}
   \label{f24}
  \vec{n} \bigr|_{r \to \infty} = n_{0} = (0,0,1)~~.
\end{equation}
Thus static configuration defines a map
\begin{equation}
   \label{f25}
 n: ~~~~~ {\mathbb S}^{3}_{\mbox{\scriptsize space}} \to
  {\mathbb S}^{2}_{\mbox{\scriptsize target}}
\end{equation}
which also can be characterized by some topological
number --- the Hopf invariant. Its origin is rather
different from the degree of map.

The characteristic feature of map
(\ref{f25})   is that the
dimension of target is smaller than that of base by
1. So the generic point
$ n $
in target has as a
preimage a whole closed line. For two different values
$ n' $ and
$ n'' $ these lines have a definite linking.
For regular configuration
$ n(x) $ such a linking does
not depend on choice of
$ n' $ and
$ n'' $. Thus an integer
appears --- the linking number of two preimages of the
map
$ n $.
This is a Hopf invariant. It is already clear
from this picture that interesting configurations are
not point--like but rather string--like.

Let us elaborate on this a little more. If our base is
2--dimensional, the field
$ n(x) $
defines the map
$ {\mathbb S}^{2} \to {\mathbb S}^{2} $
with usual degree of map.
Embedding such configurations into 3--space
one gets straight strings with infinite energy.
Natural idea is to close them
into a bounded object.
However,
a simple closure will not do.
Indeed,
the energy is proportional to the length
and simple closed string will collapse.
However,
if one twists the string and/or knots it
before closing,
there is a chance,
that such twist will increase with shortening
of the string and raise the energy.
Thus some stable configuration could appear.
These simple considerations were leading for
me in my proposal.

To construct the lagrangian, consider first
the topological current
corresponding to Hopf invariant.
Define the antisymmetric 2--tensor
\begin{equation}
 \label{f26}
H_{\mu \nu}
= ( \partial_{\mu} n \times
 \partial_{\nu} n , n )
 \, ,
\end{equation}
which is closed
\begin{equation}
 \label{f27}
\partial_{\rho} H_{\mu \nu} + \mbox{cycle} = 0
 \, ,
\end{equation}
so that there exists a vector field
$ C_{\mu} $
such that
\begin{equation}
\label{f28}
H_{\mu \nu} = \partial_{\mu} C_{\nu} -
 \partial_{\nu} C_{\mu}
 \, .
\end{equation}
Then the current
\begin{equation}
   \label{f29}
J_{\mu} = \epsilon_{\mu \nu \rho \sigma}
   H_{\nu \rho} C_{\sigma}
\end{equation}
conserves and
\begin{equation}
\label{f30}
Q = \int J_{0}~d^{3}x
\end{equation}
is a Hopf invariant.
Now the lagrangian is a natural variant
for that of Skyrme
\begin{equation}
 \label{f31}
L = a (\partial_{\mu} n)^{2} + b H_{\mu \nu}^{2}
 \, .
\end{equation}
However,
due to the nonlocality in the definition
of Hopf charge
(one is to integrate to get
$ C_{\mu} $
from
$ H_{\mu \nu} $),
the simple estimate as
(\ref{f22})
is not valid.
It was shown by Kapitansky and Vakulenko
\cite{c19},
that the following estimate is true
\begin{equation}
  \label{f32}
H_{\mbox{static}} \ge c | Q |^{3/4}
 \, .
\end{equation}
This is enough to expect the stable solutions
in sectors with
$ Q \ne 0 $.

It is natural to consider as a center
of string--like soliton the line
on which field
$ n $
has the value
opposite to the vacuum one
(\ref{f24}),
namely
\begin{equation}
  \label{f33}
n = ( 0, 0, -1 )
 \, .
\end{equation}
The simplest case is axial symmetric,
when this line is a circle.
The separation of variables
in axial symmetric case still leaves us
with two variables on the base,
so that to find a solution one is to solve
nonlinear partial differential equation
in two variables.
In the 70--ties it was beyond the possibilities
of computers.
In contrast,
for Skyrme model the spherically
symmetric
Ansatz exists,
reducing the problem to ordinary differential
equations.
This is why
the numerical realization of my hypothetical
string--like solitons waited for more than 20
years.

The first result in this direction
is due to A.~Niemi,
who agreed to learn basics of modern
programming and got the facilities
of the supercomputing center in Helsinki.
The findings as well as conjectures
were published by us in a paper in ``Nature''
with the title ``Knots and Particles''.
Needless to say,
we were to make allusions to old ideas
of Lord Kelvin
\cite{c20}.

During the last two years several
more professional groups joined in the quest
for finding string--like solitons and
quite a rich picture appears now,
see
\cite{c21},
\cite{c22},
\cite{c23}.

Thus a natural question appeared,
namely,
if it is possible to find the applications
for these results in QFT.
I shall finish my talk indicating a use of
the
$ n $--field
in description of the Yang--Mills field.
This is a programme,
which Niemi and me began to work on very
recently
\cite{c24}.

One suggestive connection between
$ n $--field
and
$ SU(2) $
Yang--Mills field is via Yang--Wu monopole.
Indeed,
if one puts
\begin{equation}
 \label{f34}
A_{\mu}^{a} = \epsilon^{abc}
 \partial_{\mu} n^{b} n^{c}
\end{equation}
then for the hedgehog configuration
\begin{equation}
 \label{f35}
n^{a} = x^{a} / r
\end{equation}
one gets exactly Yang--Wu monopole
(\ref{f15}).
The singular gauge transformation
mentioned above
is that of rotating
$ n $
from
(\ref{f35})
into constant direction like
(\ref{f24}).
This suggests
that in a particular gauge
$ n $--field
may constitute part of parameters
for the Yang--Mills field.
The field
$ n ( x ) $
is then to be interpreted as an order
parameter of the monopole condensate.

The next question is
if it is possible to complete the formula
(\ref{f34})
to include a full set of independent
field degrees of freedom.
After some deliberations Niemi and me
came to the formula
\begin{eqnarray}
\label{f36}
\vec{A}_{\mu} = C_{\mu} \vec{n}
   + \partial_{\mu} \vec{n} \times \vec{n}
   + \\
\nonumber
    \rho \partial_{\mu} \vec{n}
   + \sigma \partial_{\mu} \vec{n} \times \vec{n}
\, ,
\end{eqnarray}
where
$ C_{\mu} $
is abelian vector field and
$ \rho $
and
$ \sigma $
--- two real scalar fields.
One must consider this
$ \vec{A}_{\mu} $
as a Yang--Mills field in a partially fixed
gauge.
The gauge freedom left is the local rotation
around vector
$ \vec{n} $,
so that the infinitesimal gauge parameter
takes the form
\begin{equation}
\label{f37}
\vec{\epsilon} = {\epsilon} \vec{n}
\, .
\end{equation}
It is easy to see,
that gauge variation
\begin{equation}
\label{f38}
\delta \vec{A}_{\mu} =
    \partial_{\mu} \vec{\epsilon}
   + \vec{A}_{\mu} \times \vec{\epsilon}
\end{equation}
corresponds to
\begin{equation}
\label{f39}
\delta \vec{n} = 0
\; ; \;
\delta \vec{C}_{\mu} =
 \partial_{\mu} \vec{\epsilon}
\; ; \;
\delta \rho = - \epsilon \sigma
\; ; \;
\delta \sigma = \epsilon \rho
\; ,
\end{equation}
so that
$ C_{\mu} $
transforms as abelian gauge field and pair
$ ( \rho, \sigma) $
constitutes corresponding
charged scalar field.
In fact,
the first line in
(\ref{f36})
corresponds to a particular connection
and second line is a covariant vector field.

Connection of the first line in
(\ref{f36})
and its generalization for
$ SU(3) $
group was considered in a series of papers
of Cho
\cite{c25}.
However,
he did not consider the problem of complete
parametrization of YM field.

Let us show,
why our parametrization can be complete in
3~+~1
dimensional space--time.
Indeed,
usual counting gives 2 polarizations for every
component of the massless vector field,
thus
$ SU(2) $
Yang--Mills theory has
$ 2 \times 3 = 6 $
independent degrees of freedom.
In our parametrization we have 2 degrees of
freedom for
$ C_{\mu} $,
2 for field
$ n(x) $
and 2 for pair
$ ( \rho, \sigma) $.
Altogether it gives 6,
as needed.

Now let us briefly discuss the relevance
of the knot lagrangian
(\ref{f31})
in this new setting.
It is clear
that
$ H_{\mu \nu} $
from
(\ref{f26})
enters into the curvature of full
$ \vec{A}_{\mu} $
from
(\ref{f36}),
so that the second term in
(\ref{f31})
is already present in Yang--Mills lagrangian
$ 1 / g^{2} ~ \mbox{tr} F^{2}_{\mu \nu} $.
The first term
from
(\ref{f31})
is of course absent as well as
dimensional constant
$ a $.
However,
it can appear as an effective low energy term
after integration over all fields but
$ n(x) $.
Indeed,
it is the first relevant term of the gradient
expansion.
Thus we propose the knot lagrangian
(\ref{f31})
as an effective infrared lagrangian for QCD
with knots being candidates for QCD strings
relevant for confinement.

Returning to general parametrization
(\ref{f36})
we see that according to our proposal the degrees
of freedom of the
$ SU(2) $
Yang--Mills field separate into abelian vector field,
magnetic monopole condesate
$ n(x) $
and charged scalar field. This sounds very consonant
to 't~Hooft proposal of late 70-ties
\cite{c26},
which now goes with the name `` abelian dominance ''.
We also hope, that we have a definite proposal for
the origin of QCD strings, advertized by Polyakov
\cite{c27}.
Of course, lots of work is needed to
substantiate this rather ambitious claim.

I add here some comments on the development which
happened already after this lecture was given.

It was quite a satisfaction for Niemi and me to
realize, that our Ansatz contains the axial--symmetric
Ansatz of Witten
\cite{c28}
in his search for many--instanton solutions. Indeed,
if one puts
\begin{equation}
\label{f40}
n^{a} = \frac{x^{a}}{r}, \quad C_{k} = f(r,t) \frac{x^{k}}{r},
\quad C_{0} = g(x,t)
\end{equation}
formula
(\ref{f36})
gives exactly Witten's Ansatz. We are indebted to
W.~Kummer for a very relevant question.

Of course, to do QCD one needs
$ SU(3) $
Yang-Mills field. Some Ansatz for
$ SU(N) $
was proposed by Periwal
\cite{c29}.
However, even for
$ SU(2) $
it differs from ours. So Niemi and me recently found
a natural generalization of Ansatz
(\ref{f36})
to
$ SU(N) $
\cite{c30}.
The ingredients are
$ N-1 $
abelian vector fields,
$ \frac{N^{2}-N}{2} $
complex scalars and nonlinear field with values in
target of dimension
$ N^{2}-N $.
Altogether it gives
$ 2(N^{2}-1) $
independent degrees of freedom as is needed
for 3+1 dimensional space.

All this shows that solitons and Yang--Mills field
are more intimately connected and can go together
into QCD. I hope, that this also explains the title of
my lecture.

I stop here my propaganda for QFT in the guise of
Yang--Mills field. I understand that I do it in time
of one more decline of interest to QFT due to the
exciting development of string theory. Thus I
deliberately take the conservative stance (and
alluding to Gerard's metaphor must be considered not
as an agile rat of the end of the 60-ties, but rather as
clumsy dinosaur and should be extinct). However I do it
because I believe that QFT is still a viable
physical construction, which did not show yet all its
possibilities. If I conveyed this message to some
young people sitting here, I shall consider my
mission successful.

\end{document}